\documentclass[conference]{IEEEtran}
\IEEEoverridecommandlockouts
\usepackage{cite}
\usepackage{amsmath,amssymb,amsfonts}
\usepackage{algorithmic}
\usepackage{graphicx}
\usepackage{textcomp}
\usepackage{xcolor}
\usepackage{tabularx,colortbl,caption}

\makeatletter
\newcommand{\linebreakand}{%
  \end{@IEEEauthorhalign}
  \hfill\mbox{}\par
  \mbox{}\hfill\begin{@IEEEauthorhalign}
}
\makeatother

\def\BibTeX{{\rm B\kern-.05em{\sc i\kern-.025em b}\kern-.08em
    T\kern-.1667em\lower.7ex\hbox{E}\kern-.125emX}}

\definecolor{iqcred}{cmyk}{0.07,1,0.67,0.31}
\definecolor{iqclightred}{rgb}{246,197,206}
\definecolor{iqcblue}{cmyk}{0.72,0.15,0,0.56}
\definecolor{iqclightblue}{rgb}{178,235,235}
\definecolor{iqcblack}{cmyk}{1,1,0,0.98}
\definecolor{Watblue}{cmyk}{0.9,0.48,0,0}	
\definecolor{Watred}{cmyk}{0.07,1,0.67,0.31}	
\definecolor{Watgreen}{cmyk}{0.72,0.15,0,0.56}
\definecolor{Watyellow}{cmyk}{0,0.09,0.80,0}
\definecolor{Sciblue}{cmyk}{0.9,0.47,0,0}
\definecolor{Mathpink}{cmyk}{0.04,0.9,0,0}
\definecolor{Engpurple}{cmyk}{0.52,0.7,0,0}
\definecolor{Watgray15}{cmyk}{0,0,0,0.15}
\definecolor{Watgray50}{cmyk}{0,0,0,0.5}

\begin{document}

\title{Interactive and accessible quantum key distribution modules for introducing students to quantum science and engineering}


\author{
\IEEEauthorblockN{John M. Donohue}
\IEEEauthorblockA{\textit{Institute for Quantum Computing (IQC)} \\
\textit{University of Waterloo}\\
Waterloo, ON, Canada \\
jdonohue@uwaterloo.ca}
\and
\IEEEauthorblockN{Silas Ifeanyi}
\IEEEauthorblockA{\textit{Pearl Sullivan Engineering IDEAs Clinic, Faculty of Engineering} \\
\textit{University of Waterloo}\\
Waterloo, ON, Canada \\
silas.ifeanyi@uwaterloo.ca}
\linebreakand
\IEEEauthorblockN{Andrew Chisholm$^\dag$, Julien Côté$^\ddag$, Quazell Cunningham$^\dag$, Jamiel Nasser$^\dag$, Aaron Xayvongsa$^\dag$}
\IEEEauthorblockA{\textit{Institute for Quantum Computing and the Department of Physics and Astronomy$^\dag$ or Applied Mathematics$^\ddag$} \\
\textit{University of Waterloo}\\
Waterloo, ON, Canada}
\linebreakand \\
\IEEEauthorblockN{Evangeline Dryburgh, Fiona Dang, Gabriel Ghrayeb, Ian Jinzo Macpherson, Fareed Rasheed, Andi Zhao}
\IEEEauthorblockA{\textit{Pearl Sullivan Engineering IDEAs Clinic, Faculty of Engineering} \\
\textit{University of Waterloo}\\
Waterloo, ON, Canada}
\linebreakand \\
\IEEEauthorblockN{Simarjeet S. Saini}
\IEEEauthorblockA{\textit{Pearl Sullivan Engineering IDEAs Clinic, Faculty of Engineering} \\
\textit{University of Waterloo}\\
Waterloo, ON, Canada}
}

\maketitle

\begin{abstract}
Teaching key concepts in quantum information science and technology benefits from hands-on activities that can engage students at all levels, especially high-school and undergraduate students. In this paper, we report on an educational kit for student exploration of quantum key distribution (QKD) via optical analogies that provides an affordable and highly interactive role-playing experience for students with minimal abstraction and direct connections to applications, key quantum concepts like superposition and measurement uncertainty, and student's pre-existing knowledge like light polarization. Manual and automated versions of the kit are detailed, as well as feedback from workshops with high-school students, undergraduates, and educators.
\end{abstract}


\section{Introduction}


Quantum key distribution (QKD) is a popular entry point to QIST for learners at the high-school level and above. It allows for students to gain a hands-on intuition about key quantum concepts like measurement bases, superposition, and quantum bits~\cite{keyconcepts}, all in the context of a relatable application (cryptography) that can be explained with a narrative that allows for student role-playing. Many approaches exist for teaching core quantum concepts through QKD~\cite{folkers2026analysis}, including role-playing games with rules enforced by human referees~\cite{svozil2006staging,ferner2023card}, online simulators~\cite{kohnle2017interactive,parakh2017quasim}, and programming-based modules~\cite{sebastian2023teaching}. Hands-on, haptic approaches using analogies with classical optics reduce the level of abstract and more closely mimic the physics underlying genuine QKD implementations~\cite{utama2020hands,faerman2024design,ivory2025qcamp}, but tend to be prohibitively expensive for many educators~\cite{thorlabsedu}. This is especially true at the high-school level where they could be incredibly useful, as curriculum expectations grow to include more goals related to quantum technologies, necessitating hands-on inquiry-based activities to explore concepts from theory. Existing QKD analogy kits also tend to be slow due to manual operation, making it difficult for students to obtain long enough keys to send realistic messages in a reasonable time frame.

In this paper, we outline an affordable and easy-to-use QKD analogy kit for high-school and undergraduate quantum education built with accessible, affordable, and 3D-printed components. This version uses components very closely analogous to genuine implementations of QKD with polarization-encoded qubits, using microcontroller boards to mimic the single-photon statistics from bright classical light. We first describe the implementation of the key components of the QKD analogy for a manually operated activity. We then describe an adaptation of the kit that automates the key distribution protocol, allowing for students to experiment with larger key sizes. We also provide some feedback from students and teachers on its effectiveness and discuss extensions of the kit to other quantum education activities.

\section{QKD analogy kit components and assembly}

\begin{figure*}
    \centering
    \includegraphics[width=2\columnwidth]{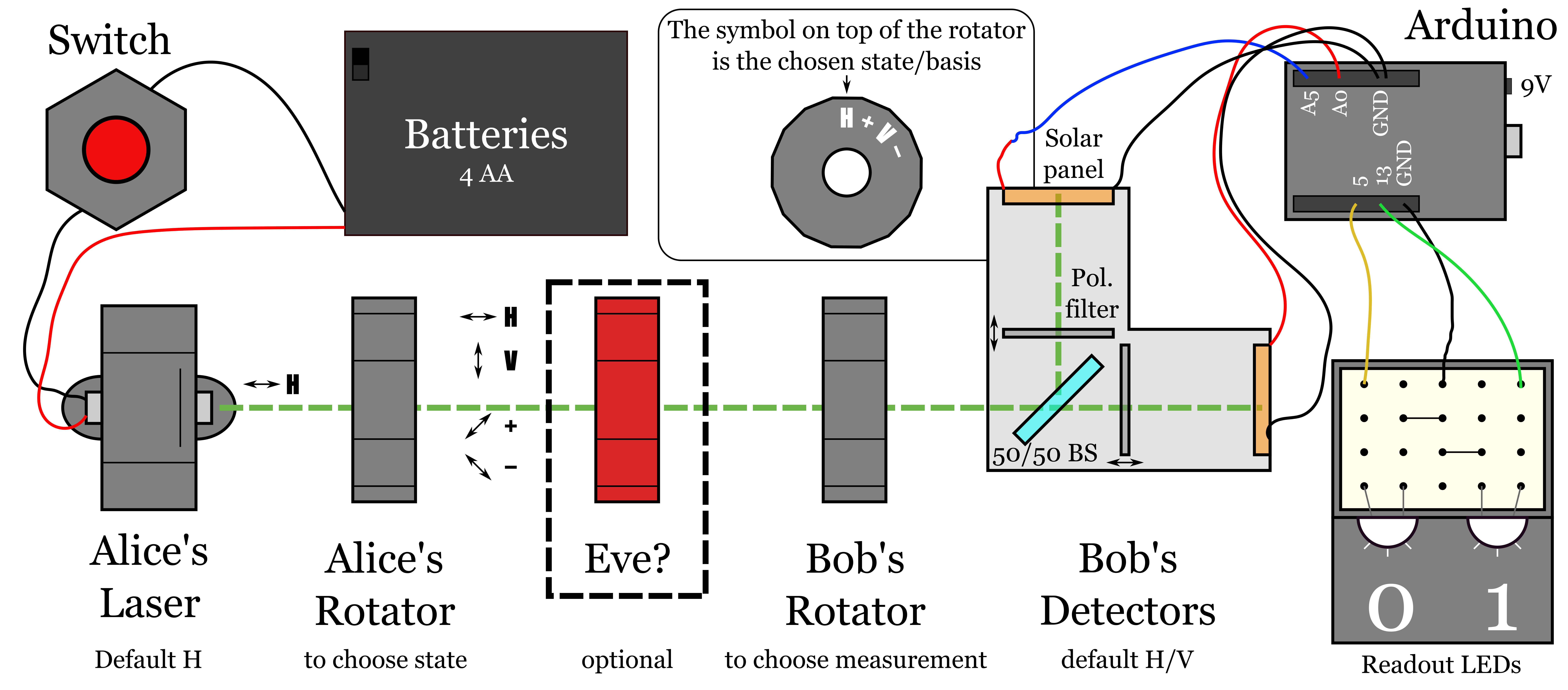}
    \caption{Top: Sketch of the hand-operated QKD kit. A laser with a push-button switch acts as Alice's photon source, by default horizontally (H) polarized. A wave-plate polarization rotator is used to choose one of the 4 relevant QKD states. Another wave-plate representing Eve can be optionally inserted after Alice to disturb the communication protocol. Bob has an identical wave-plate polarization rotator to change the measurement basis. The polarization measurement is in a light-tight box with a 50/50 beam-splitter leading to two orthogonal polarizers and solar panels to convert laser intensity to voltage. The voltages are measured with an Arduino, converting the signals to a binary output visualized with two readout LEDs.}
    \label{fig:manualkit}
\end{figure*}

\begin{figure}
    \centering
    \includegraphics[width=\columnwidth]{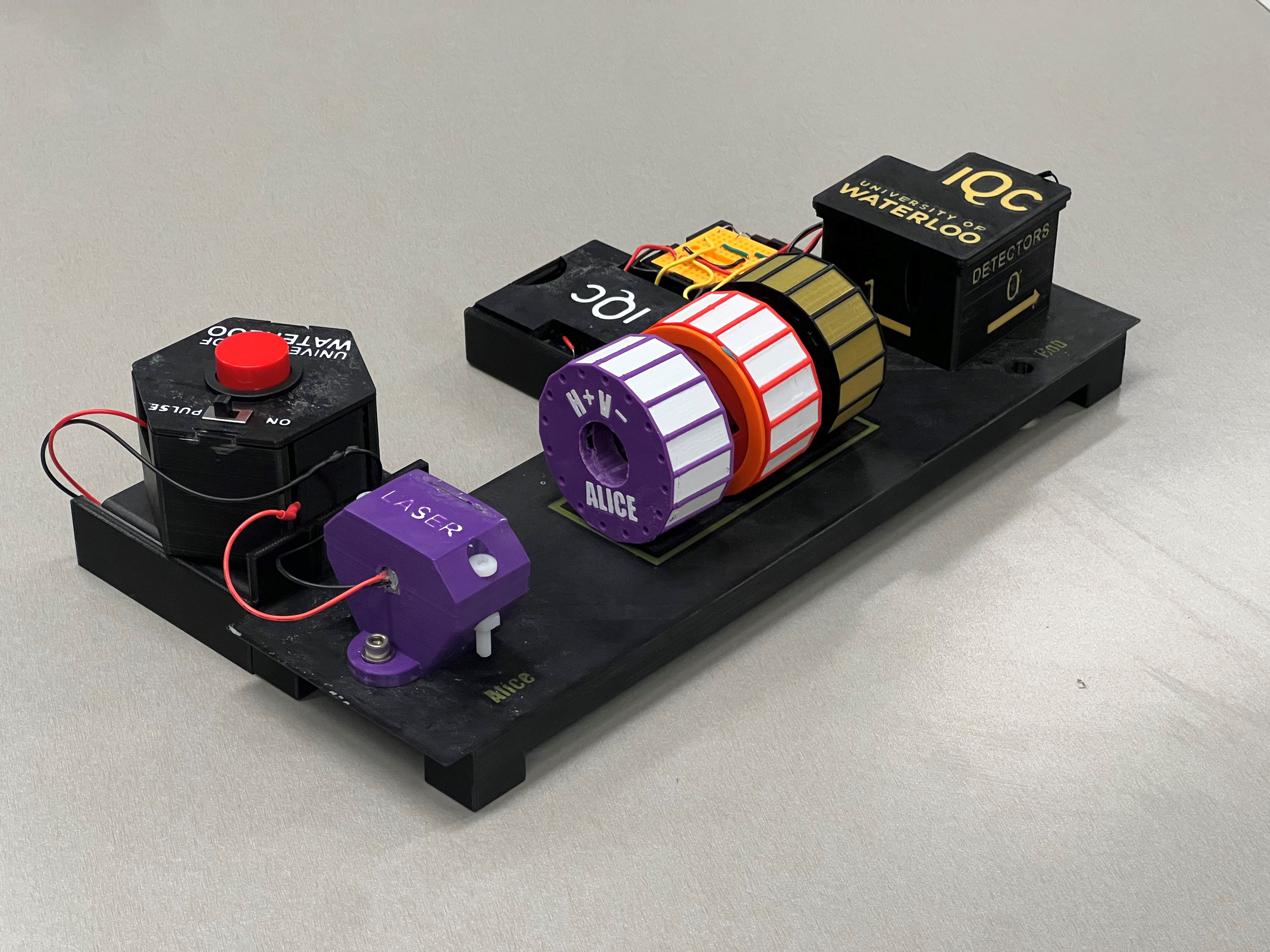}
    \includegraphics[width=\columnwidth]{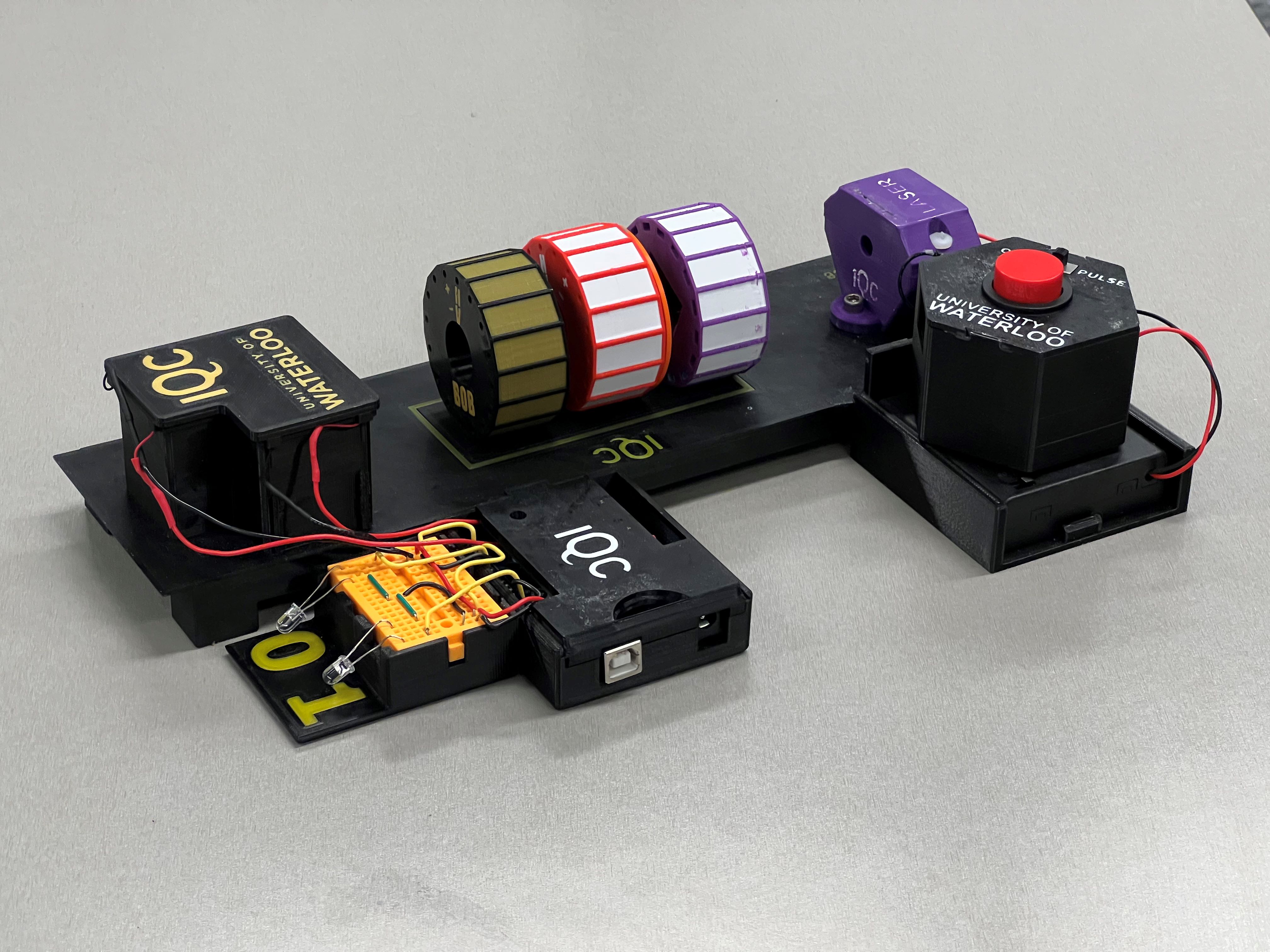}
    \caption{Photo of the QKD kit from the point of view of Alice (top) and Bob (bottom).}
    \label{fig:manualkitphoto}
\end{figure}

QKD is a protocol for generating random shared keys between two parties, traditionally labeled Alice and Bob. In the earliest QKD protocol, BB84~\cite{bennett1984quantum}, Alice sends Bob a sequence of qubits encoded in single photons, which are randomly encoded in the ``0'' or ``1'' qubit state and also randomly in one of two mutually unbiased bases; traditionally, these are the Z and X Pauli bases, which are sometimes known as the rectilinear (or ``HV'') and diagonal (or ``DA'' or ``$\pm45^\circ$'') bases for polarization qubits. Bob measures each qubit in a random one of the two bases and records his output. If Alice and Bob disagree on the basis, their results will not be correlated. However, if they agree on the basis, they should agree on the bit value as well. By publicly announcing the bases after the qubit exchange, they can confirm whether they expect to agree or not. If an eavesdropper, labeled Eve, naively measures the qubit between Alice and Bob, they will not know which basis Alice used. If they choose the wrong basis, they will change the quantum state transmitting from Alice to Bob, introducing errors where Alice and Bob should have agreed. By estimating the error rate during the key exchange, Alice and Bob can either detect the presence of an eavesdropper by noting a high error rate or place bounds on the security of their key if they observe a low error rate.

Real-life QKD experiments require quantum signals, as the no-cloning theorem prevents Eve from making a copy of Alice's qubit. However, the basic effect of intermediary measurement changing the state of a system can be noted through analogies with bright states of light, exemplified in the three-polarizer experiment~\cite{chang2025handmade}. QKD analogy kits take advantage of this parallel to allow students to experiment with QKD protocols in a hands-on manner for much lower cost, replacing genuine quantum components with bright light sources and analyzing the output signals to mimic the randomness of quantum bits. While this approach prohibits any sort of quantum-enabled secure key exchange, as they are vulnerable to attacks where the eavesdropper intercepts a small amount of the light energy Alice sends to Bob, it preserves core educational behaviour while reducing complexity and cost.

In the rest of this section, we will outline the design of a QKD analogy kit built for high-school QIST education, sketched in Fig.~\ref{fig:manualkit} and seen in Fig.~\ref{fig:manualkitphoto}. The essential elements of the kit are similar to others that have been demonstrated previously~\cite{utama2020hands} and others that are commercially as well as publicly available~\cite{thorlabsedu,haverkamp2022simple,O3Qbb84}. The primary considerations in this implementation were cost (to approach affordability for high-school educators), ease-of-use (to allow many kits to run in parallel with few facilitators), and interactivity (to allow significant hands-on experience for students). The 3D models, list of materials, and build instructions are available online in Ref.~\cite{iqcTeacherResources}, and an estimate of the cost of the kit materials can be found in Table~\ref{tab:costs}. A video showing the operation of the kit can be found in Ref.~\cite{qkdyoutube}.

\begin{table}
    \centering
    \begin{tabular}{lr}
        Item & Approx. cost (CAD) \\
        \hline
        3D filament (PLA) & \$10 \\
        Laser, 520~nm & \$52 \\
        Button & \$3 \\
        Wave-plate film & \$6 \\
        Beam-splitter & \$58 \\
        Polarizer & \$3 \\
        Solar panels & \$14 \\
        Arduino Uno & \$38 \\
        Magnets & \$12 \\
        Misc. (e.g., batteries, wires, screws) & \$10 \\
        \hline 
        Manual kit sub-total& \$205\\
        \hline 
        Servo motors & \$6 \\
        \hline
        Automated kit total & \$211 \\
    \end{tabular}
    \caption{Estimated costs for the QKD analogy kit components per kit. Values are in Canadian dollars, neglect shipping costs, assume access to 3D printers, and include some savings from buying in bulk (such as for the wave-plate film). A complete list of recommended materials and suppliers for the manual QKD kit can be found in Ref.~\cite{iqcTeacherResources}.}
    \label{tab:costs}
\end{table}

\subsection{Light source}

In QKD, qubits may be encoded into one of many photonic degrees of freedom. Polarization is a popular choice, as it is a natural binary system that is relatively straightforward to prepare and measure. Polarization is ubiquitous in QKD analogy kits as it requires no interferometric stability and can be controlled and measured with affordable, off-the-shelf components. 

In our analogy kit, we use a green laser diode powered by a 6~V battery pack as Alice's light source. For compatibility with the polarization rotators detailed in Sec.~\ref{ssec:polrot}, a green laser (wavelength near 520~nm) was necessary. We use a small piece of polarizing film to clean the polarization of the laser and initialize it close to horizontal (``H''), with an optical power of approximately 0.5~mW. The laser is mounted in a 3D-printed holder and secured at its base to point at the polarization analyzer and detectors. Between the battery pack and the laser, we introduce a momentary push-button switch which allows students to send brief pulses of laser light, in analogy with a single photon being released.

\subsection{Detectors}\label{sec:detectors}

For Bob to detect Alice's polarization state, we use a polarization analyzer consisting of a polarizing beam-splitter (PBS) and a pair of solar panels inside a 3D-printed mount. To keep costs reasonable, the PBS is built using a 50/50 beam splitter (BS) with a horizontal polarizing filter at the output of the transmitted mode and a vertical polarizer at the output of the reflected mode, each followed by a solar panel. This configuration has a minimum 50\% loss, but the light source is bright enough to compensate. To block stray background light, the mount is designed with a tight-fitting lid and a green chromatic filter is included after the polarizing film.

Below their saturation value, the solar panels produce a voltage roughly proportional to the light intensity incident upon them. The voltage across each solar panel is measured using an Arduino Uno microcontroller board, with the voltage in the horizontal (transmitted) path labeled $V_0$ and that in the vertical (reflected) path labeled $V_1$. When detecting a signal, the Arduino can cause a readout LED to flash, illuminating either a ``0'' or ``1'' which Bob reads as output. Since different panels have slightly different efficiencies (and the efficiency can vary across the length of the panel) and to avoid the need for background subtraction or significant calibration, we apply the below rules to determine if a signal was measured and what output to provide to Bob rather than assigning probabilities directly from the measured voltages:

\begin{itemize}
    \item If both panels read a voltage below a certain threshold (typically $V_t=0.8~V$), no signal is recorded and no LEDs are turned on.
    \item If only one of the two voltages is above the threshold value, the corresponding LED is turned on (``0'' if $V_0>V_t$, ``1'' if $V_1>V_t$).
    \item If both $V_0$ and $V_1$ are greater than $V_t$, but one is larger than the other by some multiplicative constant $w$ (usually 2), then the LED corresponding to the larger voltage is turned on (e.g., if $V_0>wV_1$, then the ``0'' LED will turn on).
    \item If both $V_0$ and $V_1$ are greater than $V_t$, but within a factor of $w$ of each other, a random number generator is used to light one of the two LEDs with a 50\% probability each.
\end{itemize}

The exact values of $V_t$ and $w$ may have slightly different optimal values between setups, but these rules result in a deterministic outcome when the input signal is in the same basis as the measurement and a random outcome when the bases mismatch, as expected in QKD. The measurement outcomes closely mimic the expected behaviour for a single incident photon, as only one of the LEDs may turn on at a time.


\subsection{Polarization rotators}\label{ssec:polrot}

To set the polarization state and measurement bases, Alice and Bob each have a half-wave plate. For the accuracy demanded by the algorithm detailed in Sec.~\ref{sec:detectors}, only modest accuracy in the angular setting and phase retardance is necessary. This is achieved using a thin half-wave plate (polymer retarder) film cut into 20-mm squares. Note that we were only able to find one reliable supplier for the wave-plate film, which worked well for green light around 520~nm but not for red laser light around 630~nm. By purchasing a large square of the film (300~mm by 300~mm), we can cut wave-plates for many kits with significant cost savings.

To set the angle without any fine-tuning or easy-to-break moving parts, the film is mounted in a 3D-printed 16-sided holder. Given that the light starts horizontally polarized and that we are interested in the rectilinear (horizontal and vertical, or H/V) and diagonal ($+45^\circ$ and $-45^\circ$, or $\pm45^\circ$) basis states, we only need to be able to set the wave plate to angles $0^\circ$, $22.5^\circ$, $45^\circ$, and $67.5^\circ$. We ensure that students have an easy time keeping the wave-plate at these angles by inserting a zinc bar under the kit's base and small rare-earth magnets at four positions for Alice's wave-plate (corresponding to the four polarization states she may choose from) and two positions for Bob's wave-plate (corresponding to the two basis choices).

The wave-plate mounts are designed so Alice's setting is visible to her, but not to Bob, and vice-versa. Alternative mounts can be printed that change the location and style of the labels so that the same kit can be used to explore concepts like quantum gates and projective measurements, in line with Schwinger's toys~\cite{gauvin2018playing}.

\subsection{Eavesdropper}


A key learning objective for QKD kits is the effect of eavesdropper disturbance, where the presence of an undesired agent causes errors to accumulate in the key. To implement an eavesdropper who can freely measure in the same bases as Alice and Bob, many other kits insert an entirely separate kit in between Alice and Bob, with Eve having ``fake-Alice'' and ``fake-Bob'' stations~\cite{thorlabsedu,utama2020hands}. This is in-line with the common narrative for eavesdropper strategies, but from a role-playing perspective for students, can be limited: the presence of Eve is visually obvious, and both the player assigned to be Eve and the key-exchange protocol are doomed to fail.

To keep the presence of Eve ambiguous and avoid the need for a third student to role-play Eve, we introduce Eve as a third wave-plate in a mount visually similar to Alice and Bob's. However, Eve's mount contains a quarter-wave plate instead of a half-wave plate. This wave-plate would be modeled as a state rotation rather than projective measurement, but because students have no way of measuring circular polarizations, the overall effect is the same. If Eve is set to (``measures'') the rectilinear (H/V) basis, the quarter-wave plate will not affect horizontal or vertical polarization, but will rotate the $\pm45^\circ$ states to the circular polarization states (equivalent of the Pauli $Y$ basis). If Alice and Bob are both using the $\pm45^\circ$ basis, this will result in random measurement outcomes, and the circularly polarized states are mutually unbiased to any linear polarization. Similarly, if Eve is set to the $\pm45^\circ$ basis, they will leave those states alone but rotate the horizontal and vertical states to the circular bases, producing the same effect. Only if Eve is aligned to the same basis as Alice and Bob will no errors be introduced.

This setup allows students to complete a round of key exchange without Eve to get familiar with the apparatus. Eve can then be introduced to demonstrate the measurement disturbance effect. For groups with more time, we sometimes introduce many potential eavesdroppers, some with the quarter-wave plate and others either empty or with a non-birefringent glass plate instead. Students then must determine if Eve is present or not, without relying on the visual cue of the extra wave-plate being in place.

\subsection{Free-space channel}


Real QKD networks require either carefully managed free-space channels (such as satellite links) or fibre-optic cables. The communication channel in the QKD analogy kit is simply the free space between the laser source and the detector box. The laser and detector box are securely mounted on a 3D-printed base such that they remain aligned at all times. In the region between the laser and detectors, a zinc plate is inserted under the base such that the wave-plates stay in-place and magnetically locked to the intended angles.

The same 3D-printed mounts used in this format have recesses for rare-earth magnets, allowing for an alternate style of activity where students freely place the components on a magnetic surface, such as a whiteboard. This can increase student interactivity by making laser alignment part of the process. In practice, the wiring to the batteries and Arduino make this version prone to disconnected wires and other errors.

\section{Usage with students and teachers}

Since designing the first version of the 3D-printed QKD analogy kit in 2022, IQC outreach has been able to maintain a supply of 8-12 kits to be used for outreach workshops. Since 2022, we have used the kits in 104 workshops, reaching over 3000 participants including over 2400 high-school students, 350 undergraduates, and 220 high-school teachers. We have also used it as a demonstration at over 30 festivals and science fairs. The key advantages to this kit over previously used alternatives are portability (allowing many kits to travel at once with easy in-class setup) and cost (allowing many copies of the kit to be built on a reasonable budget). The kit has also been used by partners and collaborators for quantum education projects~\cite{ivory2025qcamp}.

For high-school students in particular, the workshops have varied from one-off 60-90 minute workshops where we introduce classes to the basics of QKD after an introductory discussion on superposition and measurement bases in the context of polarized photons, to a hands-on portion of a longer-form workshop spread over many days. In the Quantum School for Young Students (QSYS), a competitive program attracting up to 45 high-school students from around the world to learn about QIST over the course of two weeks~\cite{laforest2017bringing}, the QKD demo is used as an early lab to cement ideas about probability, superposition, and measurement bases before extending those ideas to quantum computation. In 2023, students were asked to rate statements about each of the 11 hands-on lab activities held over the course of QSYS on a four-point Likert scale, with options (numerical values in brackets) of Strongly Disagree (1), Disagree (2), Agree (3), and Strongly Agree (4), which were then averaged over the 29 respondents. The QKD lab was consistently rated highly by students, with students rating the statement ``I found the experiment interesting and engaging'' with a score of 3.3, ``The experiment connected to and helped clarify the content from lectures'' with a score of 3.6 (the highest of any lab activity in QSYS), and ``The experiment was unlike activities I have seen before'' with a score of 3.5.

In a multi-day workshop for high-school teachers (Quantum for Educators)~\cite{donohue2026outreach}, teachers were provided the QKD kits to solidify their understanding after being presented a lesson plan on QKD fundamentals. In a post-workshop survey from the 2025 iteration, 81\% of teachers (17/21) rated the activity as ``Very Valuable'', with the remaining 19\% (4/21) rating it as ``Valuable.''

Activities related to QKD can be a natural fit in high-school classes, especially physics. For example, Ontario curriculum expectations in Grade 12 (senior) physics include highlighting the importance of quantum to the development of new technologies and doing hands-on inquiry to explore scientific theory related to quantum science~\cite{ontariocurrSCI}. Quantum physics is often a brushed-over topic because of a lack of time, educator confidence, and resources, despite high interest from students~\cite{richardson2024spherical}. QKD builds understanding of key concepts of quantum information science and technology on a foundation of material covered in most physics courses, such as the polarization of light, lasers, optical physics, and the photon nature of light.

\section{Automated QKD analogy kit}

\begin{figure}[b!]
    \centering
    \includegraphics[width=\columnwidth]{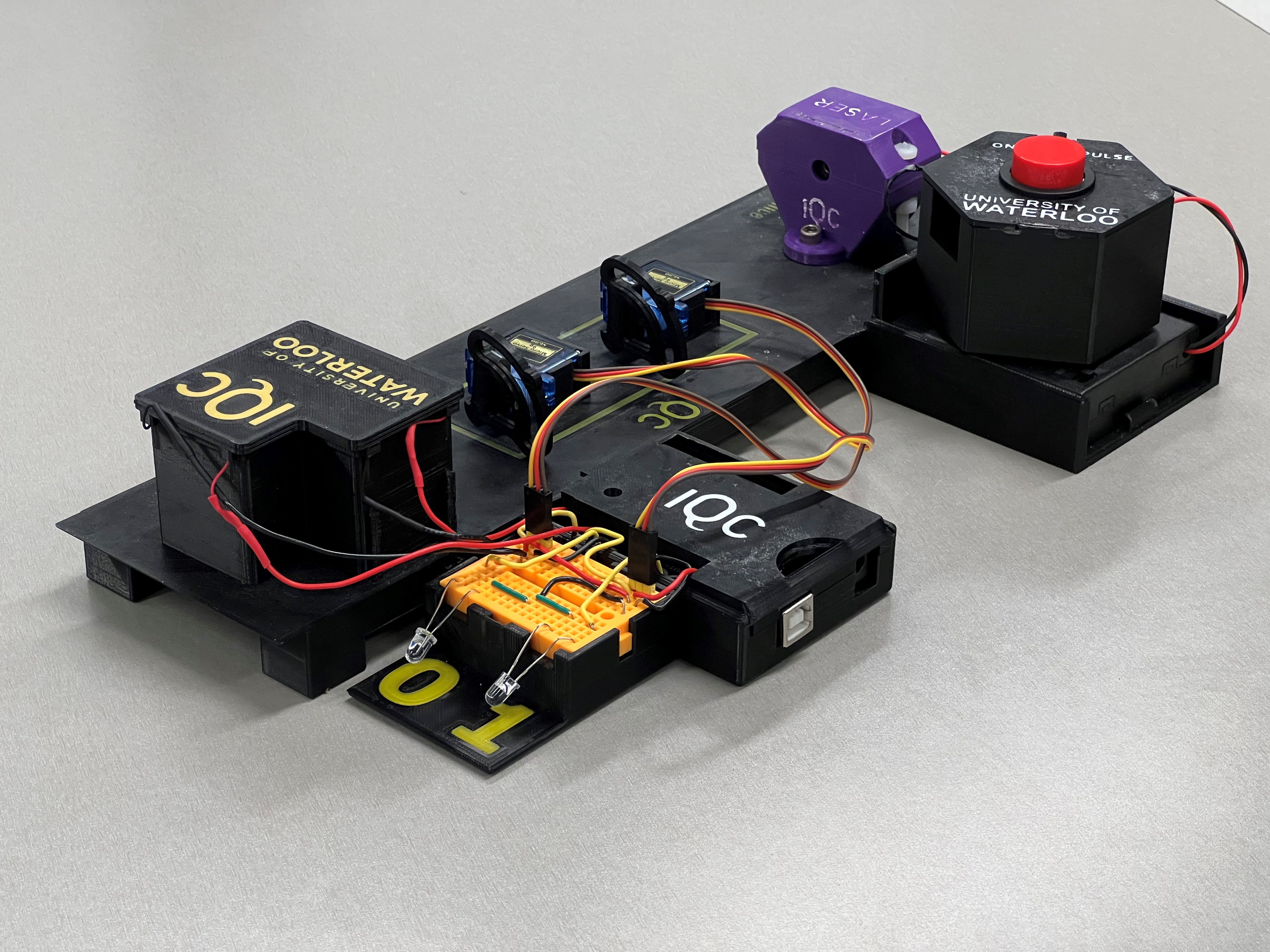}
    \includegraphics[width=\columnwidth]{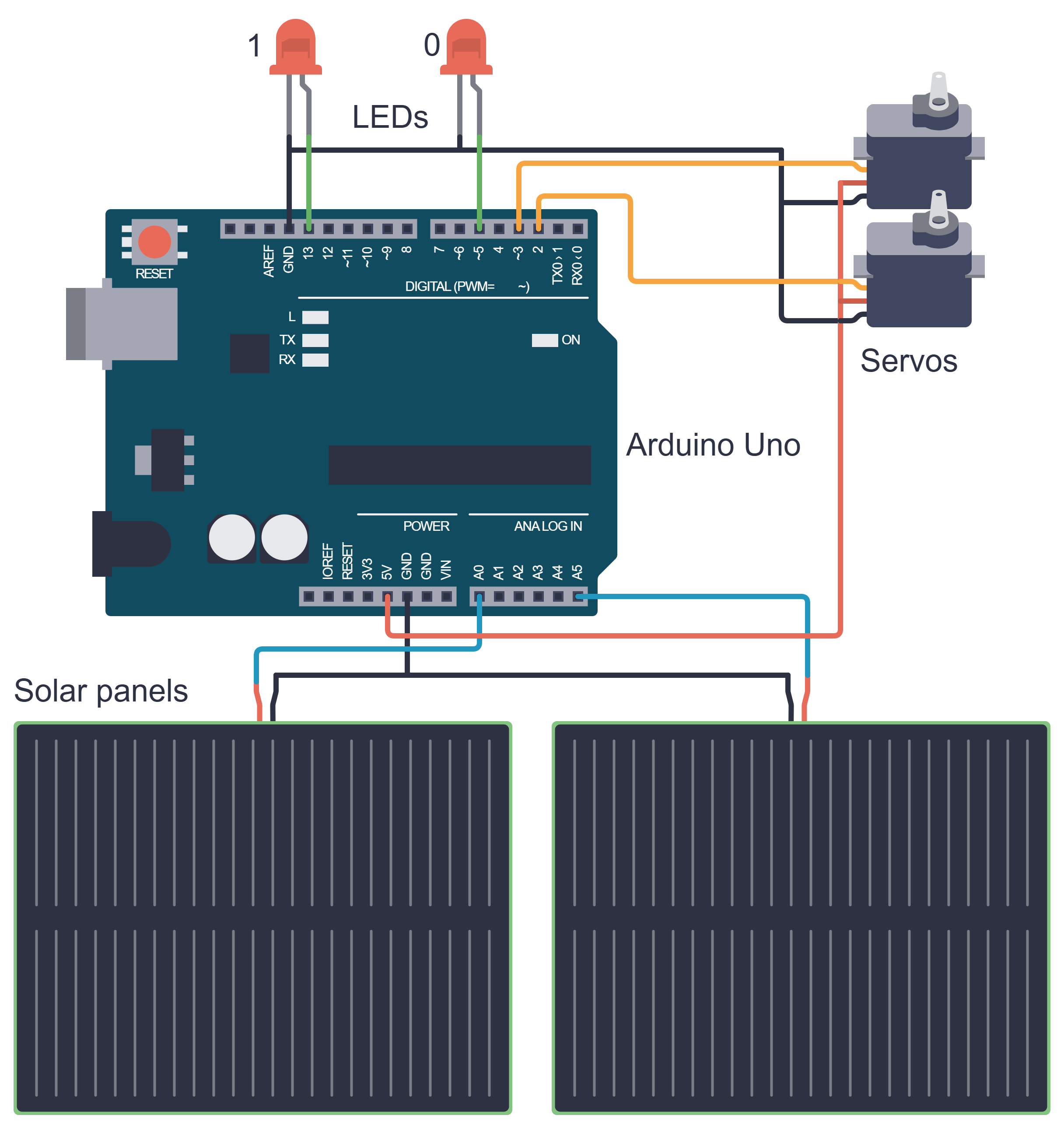}
    \caption{Top: Photograph from Bob's perspective of the automated kit, with the manually rotated wave-plates replaced by servo motors controlled by the Arduino. Bottom: wiring diagram for the automated QKD kit~\cite{circuitcanvas}.}
    \label{fig:autokit}
\end{figure}

A key limitation of the previously described QKD kit is that it is slow to operate, as students must manually rotate the wave-plates and press the button to send each bit. This makes it difficult to generate long keys in a reasonable time frame, limiting the types of activities that can be done with the kit. To address this, we have developed the automated extension to the kit seen in Fig.~\ref{fig:autokit}, which allows for the key distribution process to be run automatically by an Arduino microcontroller. This extension motorizes the wave-plates for Alice and Bob, allowing bits to be generated at a rate of 5~Hz comfortably.

\subsection{Adaptations for automation}\label{sec:motorizedwaveplates}

In order to automate the QKD protocol while maintaining affordability, we made several modifications to the assembly of the kit, most of which were centered around the addition of two servo motors connected to wave-plate mounts to change qubit states for Alice and measurement bases for Bob. These motors are controlled by the same Arduino responsible for data collection and analysis and lock in magnetically to the system in the same manner as the manual half-wave plates. Space remains for Eve's manually operated wave-plate to be inserted between the two motorized wave-plates. 

Students interact with the kit through the Arduino's integrated development environment (IDE) serial monitor, where students may enter commands to adjust wave-plate angles and read out measurements. An automatic key exchange protocol is built in which takes the number of qubits to exchange as input and generates a pseudo-random sequence for both Alice and Bob to determine states and bases, which define the sequence of settings for the servo motors. The output printed in the serial monitor contains the list of bits and bases for both Alice and Bob, with students tasked to complete the basis reconciliation and other steps of the protocol to find the key, verify its accuracy, and encrypt/decrypt a message with the remaining key. Alternative modes of operation useful for demonstration simply print the resulting key and note the presence or absence of the eavesdropper. A limitation of the current configuration is that Alice and Bob's systems are controlled by the same Arduino/computer, and separating the information meant for Alice and Bob to separate students can feel artificial.


While the servo motors have sufficient angular accuracy to position the wave plates, their ``zero'' angles were found to not be reliable between various iterations of the activity. Rather than using the physical limit switches of the servos to define zero, a homing and calibration routine is used to actively find the zero angles at the start of the activity. In this way, the system can perform an self-calibration that compensates for mechanical mounting offsets and alignment variability before automated QKD demonstrations are run. Alice's wave-plate is deliberately offset by a small amount (10-20 degrees) and Bob's is scanned over a 90-degree range in coarse steps of five degrees, with the intensity on the solar panels being logged during the process. The best value (where the light is closest to horizontally polarized) is then scanned around in one-degree steps, and the optimal value defines a new zero angle for Alice and Bob for which any future settings will account. To ensure proper behaviour, we recommend evaluating the performance with a pre-programmed test routine that quickly measures all eight combinations of Alice's states and Bob's bases. Eve's wave-plate is removed during this procedure.

\subsection{Workshops with automated QKD kits and general adoption}

The automated QKD kits have been used in workshops at the University of Waterloo as both stand-alone for a QKD workshop and as part of a longer-form, multi-day workshop, teaching students about optics and quantum phenomena broadly~\cite{ifeanyi2025modular}. Both workshops were designed to be for beginner undergraduate students, with no prior experience in quantum science or engineering required. While the stand-alone QKD workshop focuses on the basics of quantum key distribution and the BB84 protocol, the longer-form workshop complements it with other hands-on activities such as building microscope systems and imaging with scanning electron microscopes. The dual manual/automated modes of the kit allowed for students to first gain familiarity with the components and the basic optics principles such as polarization states and quantum randomness in the manual mode before transitioning to the automated mode to explore the cryptography concepts and implementations.

\subsection{Data analysis and user experience}

The automated kits have been well-received by students, with exits surveys indicating high levels of satisfaction. While not a clear indicator of interest by itself, participation at the University of Waterloo grew from 17 students to 35 students in two consecutive iterations. Interest has grown beyond the University of Waterloo as well, with the automated kits being used in workshops at other institutions. Due to the affordability and accessibility of the kit, copies were built and used in a workshop at Kwame Nkrumah University of Science and Technology (KNUST) in Ghana. In its first iteration, the workshop was attended by 20 students. The kit was further extended by graduate students at KNUST to include LED screens and explore using other cryptographic methods.

\subsection{Possible extensions}

At its core, a QKD analogy kit acts as an analogy for a one-qubit quantum device, allowing for hands-on experiments beyond quantum key distribution. By re-labeling the wave-plate mounts from the language of QKD (polarization states and bases) to the language of quantum computing (unitary operations and projection operators), other experiments are possible. Slight adaptations of the QKD kit have been used to explore the Elitzur-Vaidman bomb paradox with students~\cite{elitzur1993quantum}, where the ``bomb'' is implemented as a mystery element that either contains a polarizer or not. More generally, features of one-qubit quantum protocols can be explored, including the non-commutativity of Pauli operators and the interruptive effect of projective measurements~\cite{gauvin2018playing}. By using slightly birefringent household materials like plastic food wrap, students can see how the phase between orthogonal polarizers builds by adding more layers, ultimately producing a variable-retardance wave plate. Adding other elements like calcite crystals and optically active solutions open a wider array of quantum-computing-related experiments to perform~\cite{lefebvre2026building}. The automated kits may allow for a more detailed exploration and quantification of Malus' law than traditionally available in classes.

\section{Conclusion}

We have detailed a hands-on activity that has been used to introduce thousands of students to QIST concepts through quantum key distribution. By using existing analogies with classical light, students were able to get a hands-on experience that closely mimics the behaviour of genuine single-photon experiments. By using affordable and 3D-printed components, the activity was able to scale to classroom kit quantities for quantum outreach organizations and become affordable for interested educators. By automating the activity, longer keys can be established to increase understanding of the data manipulation needed to establish and use cryptographic keys. Future work will continue to develop supporting activities using the existing apparatus to explore more foundational quantum concepts and experiments.

\section*{Acknowledgments}

IQC is grateful for support from Innovation Science and Economic Development (ISED) Canada under the Strategic Science Fund (SSF) and the American Physical Society through the APS Innovation Fund, and for contributions to the development of the kit from H.~Gallop, S.~Daley, and A.~bin~Aamir. The IDEAs Clinic, which provided help with automation of the kits, is grateful to the group of University of Waterloo alumni at Microsoft who provided funding for this project.


\bibliography{qkd_kit}

\end{document}